\documentclass{ws-mpla}

\begin{document}

\markboth{Koji Hashimoto, Pei-Ming Ho and John E. Wang} {Birth of
closed strings and death of open strings during tachyon
condensation}

%
\catchline{}{}{}{}{}
%

\title{BIRTH OF CLOSED STRINGS AND DEATH OF OPEN STRINGS DURING TACHYON
CONDENSATION
}

\author{\footnotesize KOJI HASHIMOTO
}

\address{Institute of Physics, University of Tokyo, Komaba\\
Tokyo, 153-8902, Japan
\\
koji@hep1.c.u-tokyo.ac.jp}

\author{PEI-MING HO AND JOHN E. WANG}

\address{Department of Physics, National Taiwan University \\
Taipei 106, Taiwan\\
and National Center for Theoretical Sciences, Taiwan\\
National Taiwan University \\
Taipei 10617, Taiwan\\
pmho@phys.ntu.edu.tw, hllywd2@phys.ntu.edu.tw}

\maketitle

\pub{Received (Day Month Year)}{Revised (Day Month Year)}

\begin{abstract}
The tremendous progress achieved through the study of black holes
and branes suggests that their time dependent generalizations
called Spacelike branes (S-branes) may prove similarly useful. An
example of an established approach to S-branes is to include a
string boundary interaction and we first summarize evidence for
the death of open string degrees of freedom for the homogeneous
rolling tachyon on a decaying brane.  Then, we review how to
extract the flat S-brane worldvolumes describing the homogeneous
rolling tachyon and how large deformations correspond to creation
of lower dimensional strings and branes. These S-brane
worldvolumes are governed by S-brane actions which are on equal
footing to D-brane actions, since they are derived by imposing
conformality on the string worldsheet, as well as by analyzing
fluctuations of time dependent tachyon configurations. As further
examples we generalize previous solutions of the S-brane actions
so as to describe multiple decaying and nucleating closed
fundamental strings. Conceptually S-brane actions are therefore
different from D-brane actions and can provide a description of
time dependent strings/branes and possibly their interactions.

\end{abstract}

\ccode{PACS Nos.: 11.25.-w, 11.27.+d
}

\section{Introduction}

Time dependent solutions in string theory have recently been
explored.
One motivation is that observational cosmology is
revealing that potentially most of the energy density in our
universe is in the form of dark energy and dark matter which are
not well understood.  Further evidence confirming these
observations would mean that to truly be a fundamental theory of
gravity and gauge interactions, string theory must address and
incorporate dark energy and dark matter.

A related question whose answers will also be of interest in the
study of cosmology, is what kind of solvable time dependent
backgrounds exist in string theory and what are the phases of
matter with which string theory can supply us.  One of the major
advances along this direction has been the study of D-branes and
their decay.  Sen has proposed,\cite{Senconje} and there has now
been further evidence supporting his conjectures, that unstable
D-branes in string theory have two important properties. First, a
coincident D-brane and anti D-brane, or an unstable D-brane of
wrong dimension, both have a worldvolume open string tachyon
scalar field which is governed by a tachyon potential.  In the
case of the bosonic string, the tachyon potential has a local
maximum and a local minimum but it is unbounded from below.  The
mass of the tachyon field is of the order of the string scale.  In
the case of the superstring, the theory is invariant under $T\to
-T$ and it is believed that the tachyon potential is bounded from
below.  If the tachyon is homogeneously taken to its minimum
value, then the configuration is actually the closed string vacuum
with no D-brane and so no apparent physical open string degrees of
freedom.
Tachyon condensation in string theory is similar to the Higgs
mechanism in a field theory in that both remove the tachyon from
the spectrum; however, in string theory the infinite number of
massive open string states are also simultaneously removed. The
second important property is that an inhomogeneous tachyon
configuration can represent a lower dimensional BPS brane. For
example a kink configuration which interpolates between two vacuum
configurations on an unstable Dp-brane represents a BPS
D(p-1)-brane, and a tachyon vortex on a brane anti-brane pair is a
BPS D(p-2)-brane.

Having confirmed that it is possible to parametrically change
unstable D-branes to the closed string vacuum or lower dimensions
D-branes by varying the value of the tachyon field, the next
question one would like to answer is what role do dynamics play in
regard to these different tachyon configurations and are they
connected through physical processes. Starting from a system in a
given configuration, what phases can it dynamically evolve
between?  Not all phases are necessarily connected through
dynamical evolution since parametrically varying the tachyon
configuration may change the boundary conditions, the total energy
and other conserved quantities which are fixed during time
evolution.

In almost all interacting theories dynamics are hard to understand
and exact solutions are difficult to obtain.  Gutperle and
Strominger proposed in Ref.~\refcite{stro}, however, that string
theory (and also field theories) contains a class of interesting
Spacelike brane solutions which can be defined in three ways
including Dirichelet boundary conditions in time, time dependent
tachyon kinks on decaying branes and supergravity solutions
showing back reaction of brane decay on spacetime.  Shortly
thereafter Sen found an exact rolling tachyon\cite{roll} boundary
conformal field theory describing the homogeneous time evolution
of the open string tachyon on an unstable brane. This solution,
obtained from a Wick rotation of an array of D-branes, is also
called the full S-brane and corresponds to including the time
dependent boundary operator on the string worldsheet

\begin{equation}
S=\lambda \int_{\rm boundary}\!\!\!\!\!\!\!\!\!\!\!\!
 d\tau \;\;\cosh {X^0} \ .
\label{emo}
\end{equation}

\noindent The energy momentum tensor of this solution is
\begin{equation} T_{00}={\cal{E}}, \hspace{.3in} T_{ij}= -p(X^0)
\delta_{ij}, \hspace{.3in} T_{i0}=0 \end{equation} where the
function $p(X^0)$ decreases monotonically to zero for large values
of $X^0$.  This solution is unusual as compared to many other time
dependent solutions in that the pressure decreases to zero without
any oscillations.  Interactions on the string worldsheet boundary
correspond to turning on the open string tachyon, therefore the
rolling tachyon is a homogeneous solution in which the open string
tachyon evolves uniformly on the worldvolume of a space filling
unstable D-brane.  This solution is a bounce since it begins near
the tachyon vacuum and climbs part way up the tachyon potential
before falling back to the vacuum.  One may also interpret the
evolution to begin at time $X^0=0$.  Here the tachyon perturbation
is small and so we expect the open string degrees of freedom to be
present. Due to the presence of the tachyon potential, as time
passes the tachyon value increases and the solution rolls towards
the closed string vacuum. This dynamical open string condensation
solution therefore represents the homogeneous decay of an unstable
brane.

Given the exact solution for the homogeneous decay of the unstable
brane, a natural question to further explore is what are the
consequences of introducing perturbations and inhomogeneities in
the tachyon profile.  In the case of a usual spatially varying
kink, small perturbations of these spatial kink solutions can be
interpreted as waves on or oscillations of a D-brane.  On the
other hand as we shall review later, larger perturbations of the
D-brane can sometimes be interpreted as lower dimensional branes
and strings. In comparison, for the case of time dependent kinks
even small inhomogeneities can locally drive the tachyon field to
different vacua. If the field falls into different vacuum
configurations locally this produces a spatial kink (or other
codimension) configuration which presumably will have a
description as a stable defect. In conventional field theories the
inevitability of the production of topological defects is well
known and called the Kibble mechanism. A more precise statement is
that in a symmetry breaking phase transition the universe will
fall into different minima of the vacuum manifold since regions
which are sufficiently separated are not in casual contact.  For
example any topological defects which can exist, such as the
boundary domain walls separating the different vacua, will
inevitably form. In string theory unstable D-branes are governed
by a similar scalar field tachyon, so presumably it is possible
that their decay will also produce many lower dimensional defects
such as branes and strings. Work along this direction includes
Ref.~\refcite{Sencre,LNT}. One feature of those solutions however
is that they are singular so for example they correspond to
tachyon configurations with singularities or caustics.

In the rest of this review we will examine the fate of open string
degrees of freedom left over after the tachyon condensation of the
rolling tachyon. Then we discuss the approach of
Refs.~\refcite{Sbraneaction} and Ref.~\refcite{Sbraneevolution}
which is to formulate an action for an S-brane and search for
their classical solutions.  In particular these S-branes include
the smooth formation of topological defects. Finally we will also
provide new solutions generalizing the results of
Refs.~\refcite{Sbraneaction} and \refcite{Sbraneevolution}.

\section{Death of Open Strings}

The Sen conjectures,\cite{Senconje} especially the first property,
look self-contradictory at first sight. The very definition of a
D-brane is that it is a boundary of the string worldsheet and so
should be described using open strings.  The disappearance of a
D-brane immediately means the disappearance of the open strings,
but how do open strings describe their own disappearance? The
answer behind this strange question
might provide
an interesting new description of closed
strings
--- even theories of closed strings without D-branes might be
described solely in terms of open strings.

A useful investigative approach to this question is time-dependent
tachyon condensation, because this is a smooth way to relate a
vacuum with the unstable D-brane at the top of the tachyon
potential and the closed string vacuum at the bottom. An
interesting result\cite{roll} deduced with the exactly marginal
operator (\ref{emo}) is that the pressure of the system at late
time is vanishing. This is an indication of the disappearance of
the usual physical degrees of freedom of open strings at the end
of the tachyon condensation. Sen further proceeded to write an
``effective'' action whose classical solution reproduces this
behavior of the vanishing pressure,
\begin{equation}
 S = {\cal T} \int d^{p+1}x\;
V(T) \sqrt{1 + (\partial_\mu T)^2} \ .
\label{tacac}
\end{equation}
The potential $V(T)$ is supposed to be of run-away form for large
$T$, $V(T)\sim \exp(-cT)$ with a theory-dependent constant $c$,
and ${\cal T}$ is the tension of the unstable D-brane.  Though
this is not a low energy effective action in the usual sense in
string theory,\footnote{See Ref.~\refcite{Kutasov} for
discussions.} it has been coupled to gravity and widely applied to
cosmological setups due to its interesting kinetic term. Analyzing
this action, one can show in particular that there are no plane
wave fluctuations around the classical rolling tachyon solution of
(\ref{tacac}) at late time\cite{roll} which agrees with the
expectation that the tachyon minimum acts as the closed string
vacuum. This behavior suggests an interesting possibility: the
open string tachyon may be described by an effective action with
time dependent classical solutions around which physical
excitations disappear.

To gain insight into what mechanism comes into play during brane
decay in string theory, let us discuss this disappearance of
degrees of freedom in the ``effective'' action in detail. In
Ref.~\refcite{GHY}, it was shown that the fluctuation around the
rolling tachyon solution of (\ref{tacac}) is governed by the
so-called Carrollian contraction of the Lorentz group whose metric
is
\begin{equation}
 g_{\mu\nu} = (0,1,1,\cdots,1) \ .
\label{carroll}
\end{equation}
The time-time component of a metric denotes the speed of light,
thus this system has a vanishing speed of light and fields
governed by the metric can no longer propagate; the fluctuations
are effectively governed by a Euclidean metric. The Carrollian
limit removes the open string excitations in our case here, and
this contraction of the light cone may also occur in other systems
losing degrees of freedom.

Carrollian behavior means that a tachyon fluctuation of open
strings ceases to propagate at the late stage of the rolling
tachyon, appearing motionless on the D-brane when probed by closed
strings living in the bulk. This is the death of the open strings
suggested by the ``effective'' tachyon action (\ref{tacac}). More
precisely, the time-time component of the effective metric
approaches zero in a time-dependent manner, so the open strings
gradually lose their velocity and finally become immobile.

It is easy to generalize the above statement to also include
fluctuations of the gauge field.\cite{GHY} In string theory,
constant gauge field strength can generally be incorporated into
any effective action in the form of the open string metric. Thus
the action (\ref{tacac}) can be generalized to
\begin{equation}
 S = {\cal T} \int d^{p+1}x\;
V(T) \sqrt{-\det ( \eta_{\mu\nu} + F_{\mu\nu})}
\sqrt{1 + g_{\rm open}^{\mu\nu}\partial_\mu T\partial_\nu T} \ ,
\end{equation}
where the open string metric $g_{\rm open}^{\mu\nu}$ is given by the
field strength
\begin{equation}
 g_{\rm open}^{\mu\nu} \equiv [(\eta + F)^{-1}]^{(\mu\nu)} \ ,
\end{equation}
where the indices $\mu, \nu$ on the right hand side are
symmetrized.  At late times of the rolling tachyon, the gauge
fluctuations here are also governed by the Carrollian metric
(\ref{carroll}) further limiting the propagating degrees of
freedom.

This phenomenon arises in part from an upper limit on the
magnitude of $\dot{T}$ in the tachyon actions and contains
intriguing physics since standard field theories not coupled to
gravity do not have such bounds on the speed of the rolling of
scalar fields.\footnote{Recently, brane motion in curved gravity
background which has a similar speed limit bound\cite{ads} has
attracted interest in view of inflationary cosmology.} One
direction suggested from this phenomena is to couple it to gravity
and see the consequences relevant to cosmology and
braneworlds.\cite{cosmo} As we stated earlier, however, one might
object that the action (\ref{tacac}) doesn't stand on a basis firm
enough to trust any string theoretical calculation based on it. To
properly answer such questions, one should use a tachyon action
which is derived from string theory first principles.  The tachyon
however is not massless but has a string scale mass squared so all
field theories for tachyons are not properly valid low energy
effective theories.
Nevertheless, one can write an off-shell lagrangian for a limited
profile of the tachyon field by using a boundary string field
theory (BSFT).\cite{bsft} For a tachyon linear in a spacetime
coordinate, the lagrangian is written as
\begin{equation}
 S = {\cal T}\int d^{p+1}x \; e^{-\frac{T^2}{4}}
\frac{y 4^y \Gamma(y)^2}{2 \Gamma(2y)}
\ , \quad y \equiv (\partial_\mu T)^2 \ .
\label{bsftac}
\end{equation}
This action played a main role in the verification of the second
part of the Sen's conjectures, that is, reproduction of D-brane
tensions from the tachyon topological defects.\cite{ver} Although
this action is valid only for the linear tachyon profiles, one
might try to do an analysis similar to that performed for
(\ref{tacac}) by using the time dependent solution found in
Ref.~\refcite{lint}. Surprisingly, the result of the fluctuation
analysis is the same
--- the fluctuations of the tachyon field and the gauge field are
subject to the Carrollian metric (\ref{carroll}). Therefore it is
expected that the Carrollian contraction is a reasonably accurate
description at the late stage of the rolling tachyon resulting in
the freezing of some low energy degrees of freedom.

The above analysis however is insufficient since it covers only
tachyon and gauge fluctuations, therefore overlooking the possible
interactions with other string states. Furthermore, it is not
clear if we can really trust the result, because we used the BSFT
action (\ref{bsftac}) or the ``effective'' action (\ref{tacac})
beyond its limit of validity. To overcome these difficulties, one
of the authors together with Terashima developed a technique in
BSFT to calculate mass-shell conditions of any string excitation
on various linear tachyon background profiles.\cite{HT2} Using
this technique, it was shown in Ref.~\refcite{HT1} that in fact
all the open string excitations are subject to the Carrollian
metric in the rolling tachyon background.  This then shows that
all open string states disappear during tachyon condensation.

The closed string vacuum is a non-perturbative vacuum from the
viewpoint of open string theory containing off shell physics,
which are naturally explored using schemes such as string field
theories.  In the case of cubic string field theory\cite{CSFT} the
disappearance of the open strings has been described in a
different manner, see Ref.~\refcite{CSFTv}. In that approach the
static vacuum exhibits trivial cohomology describing the
disappearance of the physical states around it, while the rolling
tachyon solution is not well understood.

\section{Birth of Closed Strings}

In addition to the death of open strings, tachyon condensation is
also accompanied by the birth of closed strings. This can be
(partially) understood from the viewpoint of unstable D-branes as
follows. The rolling tachyon creates electromagnetic fluctuations
from the time-varying vacuum, in a way analogous to the creation
of cosmological density perturbations during inflation. A scalar
field $\phi$ living in an expanding universe has the kinetic term
$a^{-2}(t) \dot{\phi}^2$ in its Lagrangian density, where $a(t)$
is the scale factor. Due to the factor $a(t)^{-2}$, the canonical
quantization is time dependent, and the induced Bogoliubov
transformation is responsible for the creation of particles.
Similarly, since the conjugate momentum for the gauge potential
$A$ is $D \simeq V(T) \dot{A}/\sqrt{1-\dot{T}^2}$ for small $E$,
which is also time-dependent due to the rolling tachyon,
fluctuations of the electric and magnetic fields are created. It
is natural to identify closed strings with electric fluxes (and
D-strings with magnetic fluxes) on the unstable D-brane upon
tachyon condensation. What is needed is a mechanism which confines
electric fluxes into infinitesimally thin bundles, which can be
identified with strings. Assuming this
confinement,\cite{confinement} the closedness of the string is
ensured by the conservation of electric flux.

The confinement mechanism can be heuristically understood as
a result of electric flux conservation and energy minimization.
The Lagrangian density of an unstable D-brane is of the form
\begin{equation}
{\cal L} = - \sqrt{1-E^2} \tilde{\cal L}(T, z), \quad
\mbox{where} \quad z = - \frac{\dot{T}^2}{1-E^2}, \quad E = \dot{A},
\end{equation}
where we assumed spatial homogeneity for simplicity.
Without specifying the explicit expression for $\tilde{\cal L}$,
the Hamiltonian and the electric flux are
\begin{equation}
H = \int \frac{D}{E}, \quad \Phi = \int D,
\quad \mbox{where} \quad D = \frac{\partial\cal L}{\partial E}
= \frac{E}{\sqrt{1-E^2}}\left( \tilde{\cal L} - 2z \frac{\partial \tilde{\cal L}}{\partial z} \right).
\end{equation}
Apparently, in order to minimize the energy for given flux $\Phi$,
one needs $|E|$ to be close to its maximal value $1$. However,
$|E| \rightarrow 1$ implies that $|D| \rightarrow \infty$. For
given $\Phi$, this means that the flux is localized in a thin
strip of infinitesimal width.

It is remarkable that the unstable D-brane worldvolume theory can
describe the process of its decaying into closed strings, but it
is even more remarkable that, as we will see, this process is
captured by a class of solutions of the S-brane action, which only
describes the $T=0$ subspace on the unstable D-brane worldvolume.

The analysis in the previous section of the death of open strings
holds in the presence of approximately uniform brane decay.
Previously we mentioned the possible role inhomogeneities in
defect formation during tachyon condensation.  Our approach is to
include inhomogeneities in the tachyon decay process and find a
controlled way to describe their effects on the rolling tachyon
background.

Let us begin with some qualitative remarks on defect formation for
a simple model, a scalar field $\Phi$ governed by a potential with
a local maximum between two local minima.  In a local neighborhood
$\Phi$ is perched near the maximum and then time evolves into the
two local minima forming a defect. Important features of the
process can be understood just by following the regions,
$\cal{S}$, where the scalar field is at the local maximum.  To
understand this point it helps to first consider the case where
$\Phi$ is sufficiently homogeneous so no defects are generated in
the local neighborhood.  $\cal{S}$ in this case is a spacelike
worldvolume giving information only about when the field $\Phi$
was at the maximum. Now introduce small inhomogeneities which lead
to the creation of defects.  For $\cal{S}$ these perturbations are
initially small, changing the spacelike trajectory only slightly.
As the perturbation grows larger and the defect is created,
however, $\cal{S}$ is deformed to a timelike worldvolume along the
trajectory of the defect. This is a partial deformation of
$\cal{S}$ in the time direction of the otherwise spacelike
worldvolume. While the spacelike regions of $\cal{S}$ do not
describe any defects, the timelike regions can be interpreted as
solitonic defects. Therefore a general feature of the time
dependent symmetry breaking process and the formation of defects,
is that existence of worldvolumes $\cal{S}$ with regions of both
spacelike and timelike signatures.

Our viewpoint will be to analyze in string theory the region
$\cal{S}$ where the open string tachyon is at the maximum of its
potential, $T=0$.  In the case of a static kink configuration of the
tachyon theory, such
a region is simply the location of the (lower dimensional) BPS
D-brane.  BPS D-branes obey the well-known DBI (Dirac-Born-Infeld)
action in the limit where
backreaction can be ignored. This comes from the fact that to
preserve conformal invariance of the string action in the presence
of boundaries, the boundaries are subject to equations of motion
which can be derived from the DBI action. In the so called Monge
(or static) gauge, the DBI action can be written with the excitation
of a scalar field associated to fluctuations in directions
transverse to a flat D-brane worldvolume hypersurface

\begin{equation} S={\cal T}_{{\rm D}p}\int d^{p+1} x
\sqrt{-\textup{det}(\eta_{\mu\nu} +
\partial_\mu \phi \partial_\nu \phi +
F_{\mu\nu})}={\cal T}_{{\rm D}p} \int d^{p+1} x
\sqrt{-\textup{det}(g+F)} \ . \end{equation} The Dp-brane is
parametrized by $p+1$ worldvolume coordinates $x_i$ ($i=0, 1,
\cdots, p$), ${\cal T}_{{\rm D}p}$ is the brane tension, the gauge
field strength is $F_{\mu\nu}$ and the fluctuation of the brane in
a transverse spatial direction is specified by the scalar field
$\phi$.  Due to the curving of the D-brane worldvolume along the
$\phi$ direction, the induced metric on the brane is
$g_{\mu\nu}=\eta_{\mu\nu}+\partial_\mu \phi
\partial_\nu \phi$. In the limit of a flat D-brane with no
excitations, $\phi=0$, this clearly gives a well defined action
for time-like extended D-branes.

\begin{figure}[tp]
\begin{minipage}{55mm}
\begin{center}
\includegraphics[width=6cm]{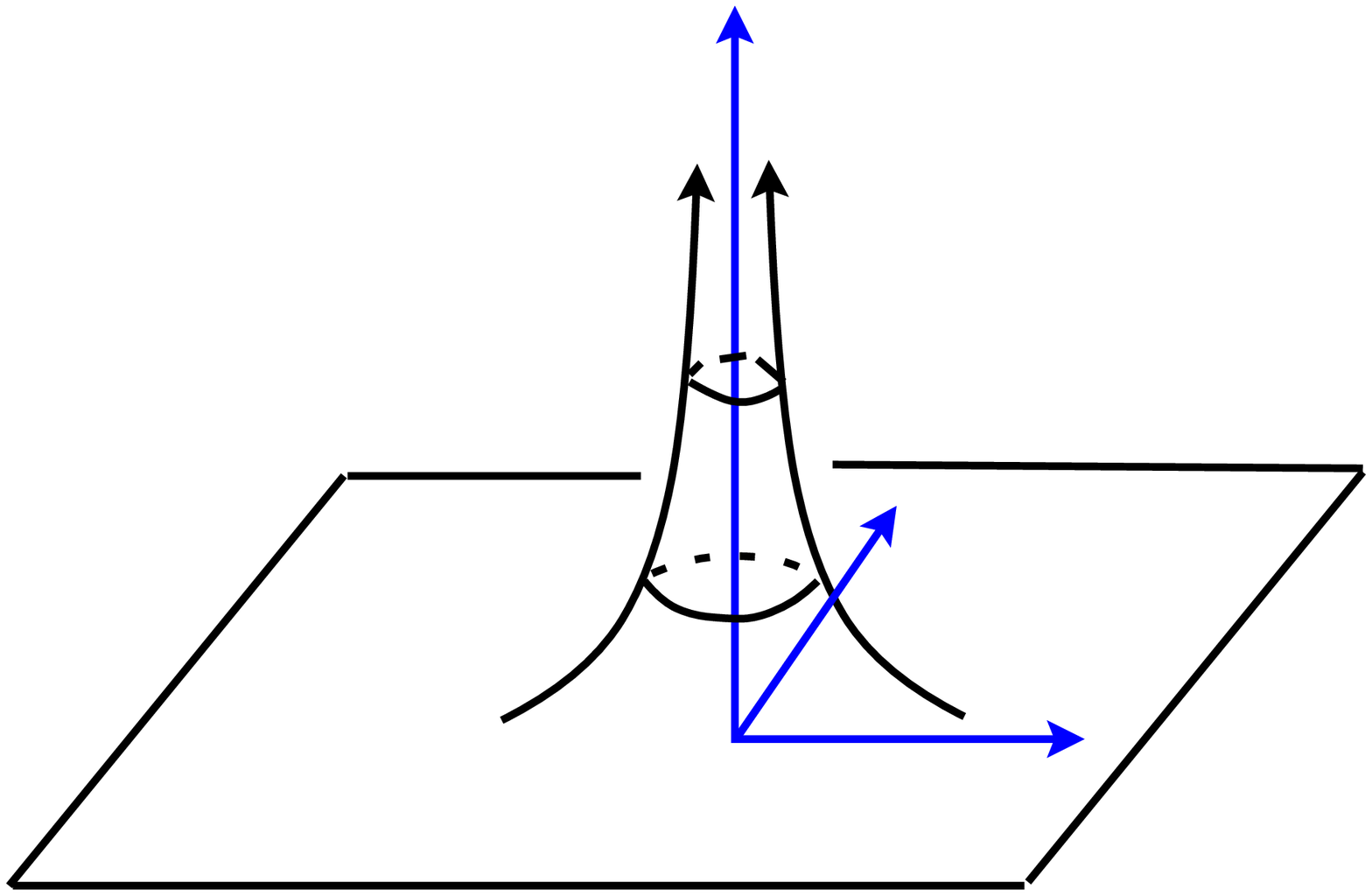}
\put(-70,100){$\phi(r)$}
\put(-50,40){$r$}
\caption{Intersection of a D-brane and fundamental string.}
\label{string-Bion}
\end{center}
\end{minipage}
\hspace*{10mm}
\begin{minipage}{55mm}
\begin{center}
\includegraphics[width=6cm]{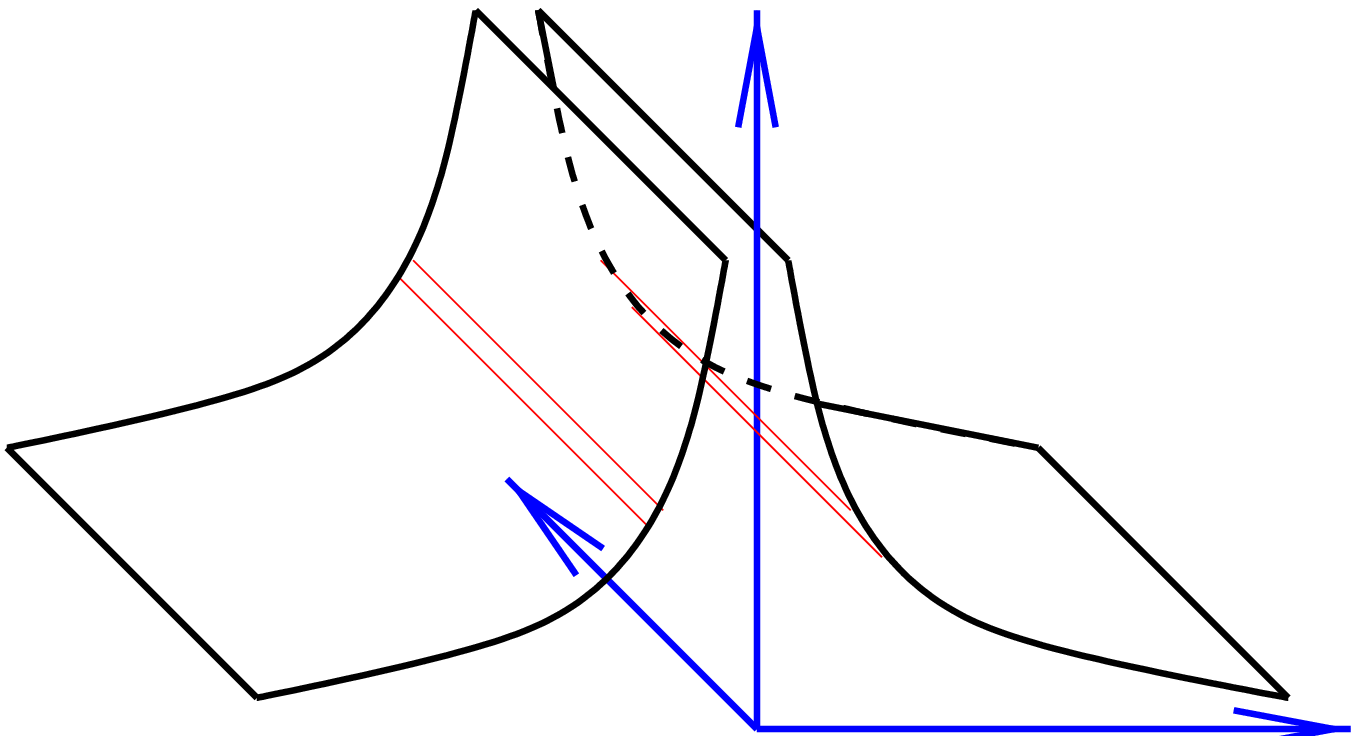}
\put(-70,90){$X^0(r)$} \put(-118,38){$x_{p+1}$} \put(-10,10){$r$}
\caption{Intersection of an S-brane and fundamental string. This
shows the formation of a fundamental string during time-dependent
 tachyon condensation.}
\label{spike}
\end{center}
\end{minipage}
\end{figure}

Solutions to the DBI action include the very interesting
BIon\cite{CM,gibbonsBI}
\begin{equation} \phi=A_{0}=\frac{b_p}{r^{p-2}} \label{bion-eq} \end{equation}
which is a perpendicular intersection of a D-brane and a
fundamental string. See Fig.~\ref{string-Bion}.  Here $A_0$ is the
Coulomb potential, $b_p$ is the electric charge,
and $r$ is the radial coordinate
$r \equiv \sqrt{x_1^2+...+x_p^2}$.
This solution is an example of
how open string degrees of freedom on the D-brane can describe a
closed string.

Primarily only timelike D-branes have been examined, but with the
introduction of S-branes, the relevance of also constructing
classical solutions of the equation of motion which have spacelike
worldvolumes has become more clear.  It is possible to show that
spacelike solutions to the equations of motion exist and in fact
are classical solutions of the Euclidean DBI action

\begin{equation} S=S_0 \int d^{p+1} x \sqrt{\textup{det}(\delta_{\mu\nu}-
\partial_\mu X^0
\partial_\nu X^0+F_{\mu\nu})}=S_0  \int d^{p+1} x \sqrt{\textup{det}(g+F)}
\end{equation}

\noindent where $S_0$ is a normalization constant.  The
worldvolume is parametrized by the coordinates $x^i$,
$i=1,...,p+1$ and so now transverse excitations can be along the
timelike direction $X^0$. In the limit of no fluctuations,
$X^0=0$, the worldvolume is spacelike so these are
Spacelike-branes.

While Euclidean solutions stand on a firm basis and arise very
naturally, they typically are considered to be associated to
quantum effects such as instantons.\footnote{The euclideanized DBI
has been used for various instanton effects, especially in
compactification of target space, as D-branes wrapping various
cycles in Calabi-Yau manifolds. These are relevant to what is
called worldsheet instantons. }  As a result of the introduction
of the notion of S-branes, however, it is now possible to
interpret some of these Euclidean solutions as classical solutions
in ordinary spacetime instead of as instantons.  The important
difference is to look at Euclidean solutions embedded in
background spacetimes which are timelike and with timelike
excitations.  In fact it has been shown that the Euclidean DBI
action naturally arises during time dependent tachyon condensation
and has been derived from the boundary string field theory and
related tachyon actions as discussed in
Ref.~\refcite{Sbraneaction}. To understand how Euclidean solutions
are related to classical solutions, let us recall the homogeneous
time dependent tachyon kinks.  Although naively a flat S-brane is
an object which is moving with infinite velocity, by examining its
associated rolling tachyon there is in fact no violation of
causality or even energy transfer associated with this solution.
This Euclidean hypersurface merely corresponds to the homogeneous
time evolution of the tachyon from its $T=0$ unstable maximum
(which one may call a ``time dependent tachyon kink'').

If the gauge field is turned off, then spacelike solutions of the
S-brane equation of motions in Lorentzian backgrounds are called
maximal surfaces in the literature since variations of the
embedding generically reduce the worldvolume. We will be
interested in solutions which change their worldvolume signature
with respect to the background geometry and so these will also be
timelike solutions to the S-brane equations of motion and have
been less studied.

Solutions which change signature are allowed in the absence of
gauge fields and the equations of motion are satisfied across the
transition point.\cite{gibbons-ishibashi} These solutions however
go through a zero in the action around which higher order effects
could be important.  In contrast, by turning on a gauge field on
the S-brane, it is possible to obtain solutions to the equations
of motion which smoothly change signature and always have finite
action.  On the worldvolume of a D-brane, gauge fields can be
interpreted as lower dimensional D-branes and strings.
Gauge fields on an S-brane will have a similar interpretation as
being branes and strings.  We next discuss S-brane solutions which
exist for long periods of time and describe time dependent defect
formation associated to the presence of gauge fields.

Intersecting S-branes solutions analogous to the BIon solutions
were found in Ref.~\refcite{Sbraneaction}. For the cases $p\geq 3$
the solution is

\begin{equation} X^0= A_{p+1}=\frac{c_p}{r^{p-2}}\end{equation}

\noindent where $A_{p+1}$ is the electric gauge field along the
$p+1$-direction whose strength is parameterized by $c_p$.  Here
$r=\sqrt{(x_1)^2 + \cdots + (x_p)^2}$ as before, but note that
$x_{p+1}$ is not included in the definition. See Fig.~\ref{spike}.
The induced metric on the S-brane is

\begin{equation} ds^2=dx^2_{p+1}+
\left(1-\frac{c_p^2}{r^{2p-2}}\right) dr^2 + r^2
d\Omega^2_{p-1} \end{equation}

\noindent which is spacelike for $r>(c_p)^{1/(p-1)}$ (or
equivalently $0<X^0<(c_p)^{1/(p-1)}$) and timelike for smaller
radial values (later time). This metric clearly describes a
cylindrical tube collapsing into a thin line. When the radius is
small, this corresponds to being at late time and far from the
original appearance of the S-brane surface. Although the metric
changes signature we stress that the action is finite and smooth
across the signature change. The induced electric field strength
of the solution, $F_{X^0 x^{p+1}}=1$ in the unit $2\pi\alpha'=1$,
is the critical electric field value. In fact the tension
generated by this electric field is exactly the fundamental string
tension!  As in the case of the BIon, this energy condition is a
result of charge quantization along the S-brane worldvolume.  This
S-brane therefore also has an interpretation as an example of the
confinement of electric flux into a fundamental string.  While
previous BIon solutions were intersections of D-branes and
fundamental strings, these new solutions can properly be called
intersections of S-branes with fundamental strings.

From our discussion of the Carrollian limit, we can provide an
intuitive explanation for the existence of solutions which
smoothly change worldvolume signature.  As defined in the context
of the worldvolume action, the spacelike brane is a solution which
is spacelike relative to the open string metric, $g^{\rm open}_{
\mu \nu}$ and not the background closed string metric $\eta_{ \mu
\nu}$. In the absence of a gauge field, both metrics are equal so
a spacelike brane can be spacelike relative to the closed string
metric.  If the gauge fields are turned on however, the open and
closed string metrics are not equal.  This then allows for the
existence of an classical S-brane which is spacelike relative to
the open string metric and yet timelike relative to the closed
string metric.

More precisely, the light-cone defined by the open string metric
lies inside the light-cone of the closed string
metric;\cite{GibbonsHerdeiro} the velocity of light on the brane
is smaller than that in the bulk by the ratio $\sqrt{1-E^2}$.
(When $E^2 = 1$, like in the Carrollian limit, open string degrees
of freedom stop propagating along the directions transverse to the
electric field.) It is therefore possible for a superluminal
fluctuation on the brane to obey the causality constraints in the
bulk. For such configurations, the S-brane Lagrangian is real (and
the usual D-brane Lagrangian is imaginary).

When the tachyon condenses homogeneously, the usual open string
degrees of freedom die due to the closing of the open string light
cone in the Carrollian limit. In contrast, the closing of the
light cone effects these S-brane solutions differently, so it
again appears that closed strings have a description in terms of
open string degrees of freedom on the brane. Therefore while open
string degrees of freedom may die, under small perturbations
closed string degrees of freedom may be born during the tachyon
condensation process!

\section{New solutions}

There are two well known extensions of the D-brane BIon solutions
given above.  BIons are supersymmetric as is typical for harmonic
function type solutions.  It is therefore possible to construct an
arbitrary configuration of fundamental strings all in parallel to
one another and perpendicularly intersecting the D-brane
worldvolume. Due to the fact that the solutions are
supersymmetric, an arbitrary configuration of strings can be
represented by simply summing the above harmonic solutions
together for any number of individual strings.

\begin{figure}[bp]
\begin{minipage}{60mm}
\begin{center}
\includegraphics[width=6cm]{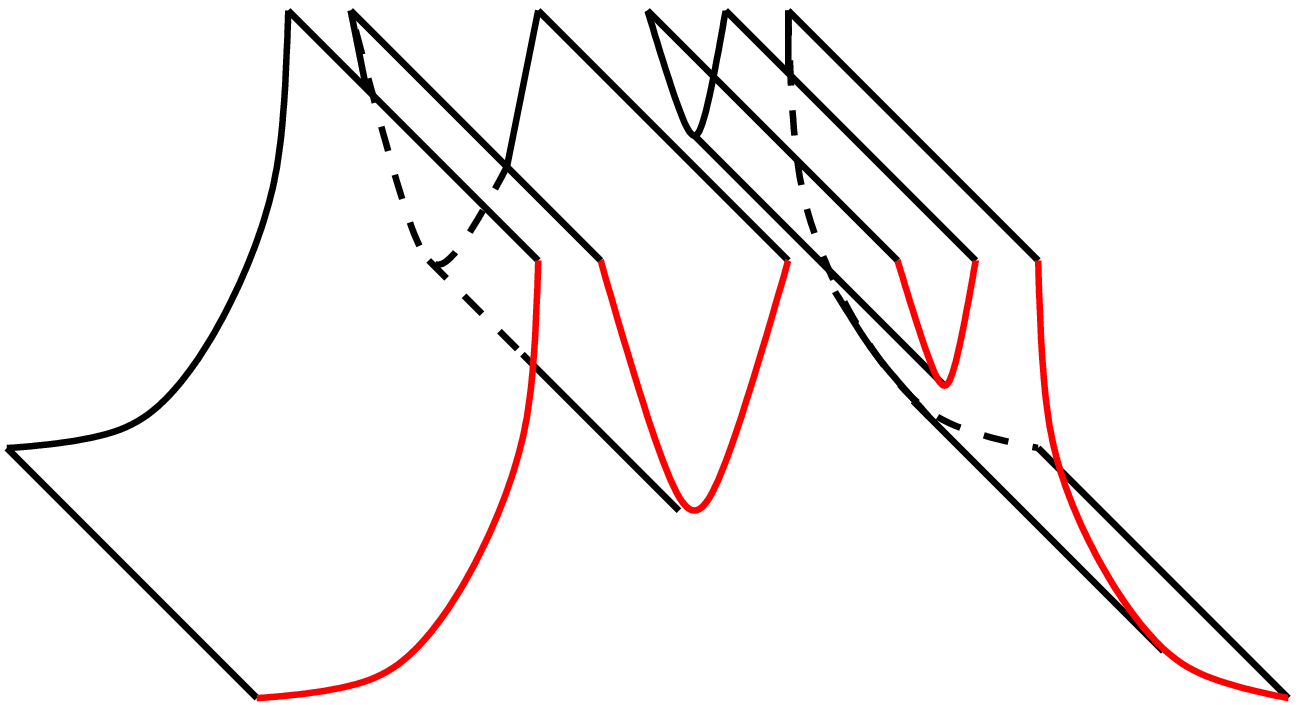}
\caption{Several closed strings created during the rolling tachyon
 process. While the S-brane worldvolume looks disconnected
  in the figure it is actually connected; this figure does not show the angular
  directions.}
\label{multispike}
\end{center}
\end{minipage}
\hspace*{3mm}
\begin{minipage}{50mm}
\begin{center}
\includegraphics[width=6cm]{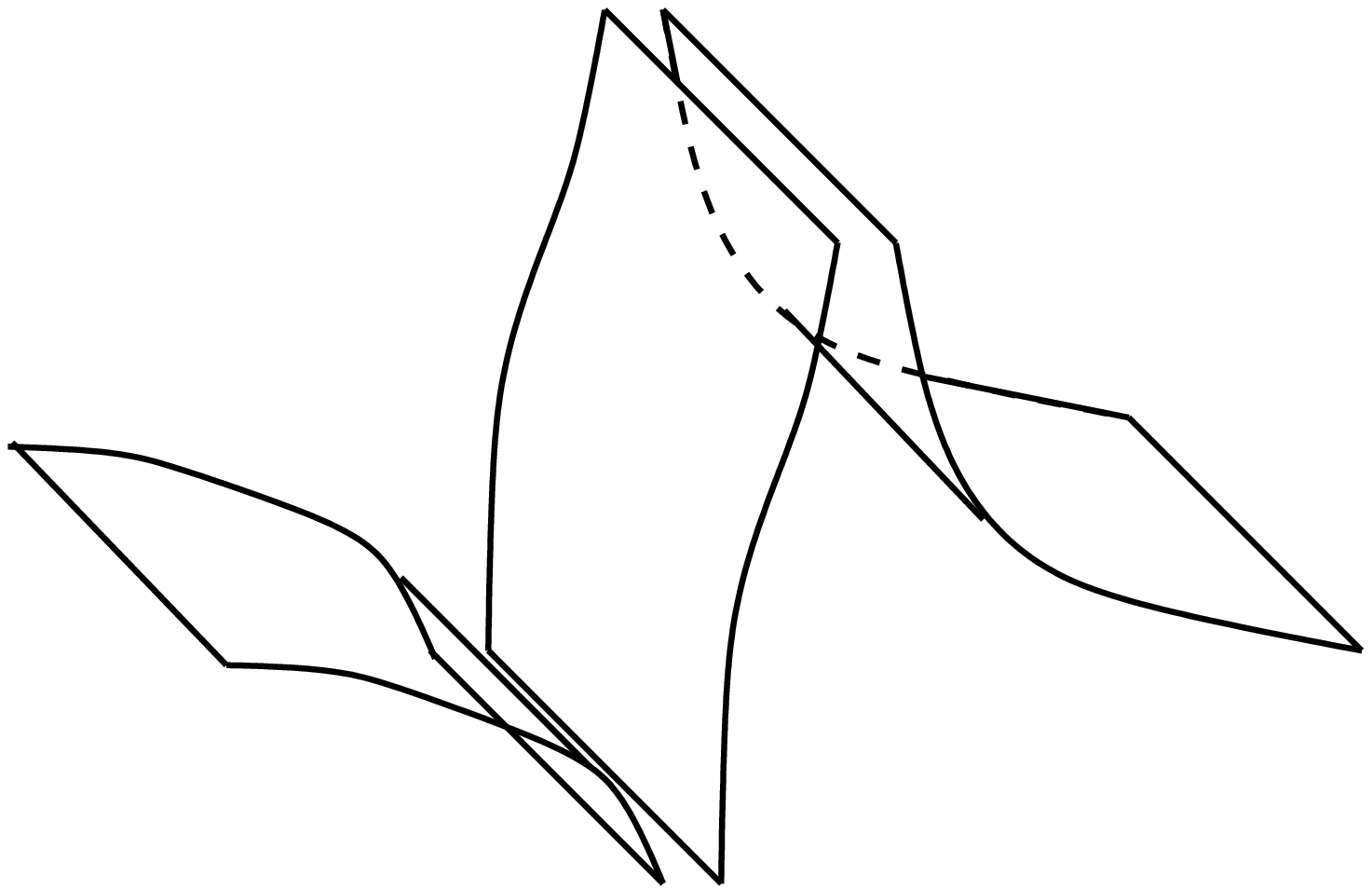}
\caption{A closed string also present in the past with $c_p<0$.}
\label{spike2}
\end{center}
\end{minipage}
\end{figure}

Surprisingly even though these intersecting S-brane configuration
are not supersymmetric, it is again possible to construct a
multiple spike solution by adding the above harmonic functions
together to obtain

\begin{equation} X^0=A_{p+1}= \sum_a \frac{c^a_p}{[(x_1-x_1^a)^2
+....+(x_p-x_p^a)^2]^{(p-2)/2}} \ .\end{equation}

\noindent The strings are labeled by the integer $a$.  This
solution has an interesting spacetime interpretation. It begins as
an approximately cylindrical tube similar to the formation of a
single string. As time evolves however local bumps and structures
develop on the tube until eventually the S-brane separates into
many disconnected tubes (see Fig.~\ref{multispike}). From a
spatial point of view this would appear to be a change in topology
proceeding in the sequence $R\times S^{p-1}\rightarrow R\times
(S^{p-1}\oplus S^{p-1}) \rightarrow \cdot \cdot \cdot \rightarrow
R\times \Sigma_i(S^{p-1})_i$. Although the cylinders become
spatially disconnected, from the spacetime point of view they are
all described by a single smooth S-brane embedding.

When some of the parameters $c_p^a$ become negative, the
corresponding spikes in the time direction are oriented towards
the past. This is the situation where we have several closed
strings also in the past and they decay once before finally
forming other closed strings in the future (see
Fig.~\ref{spike2}).

\begin{figure}[bp]
\begin{center}
\includegraphics[width=6cm]{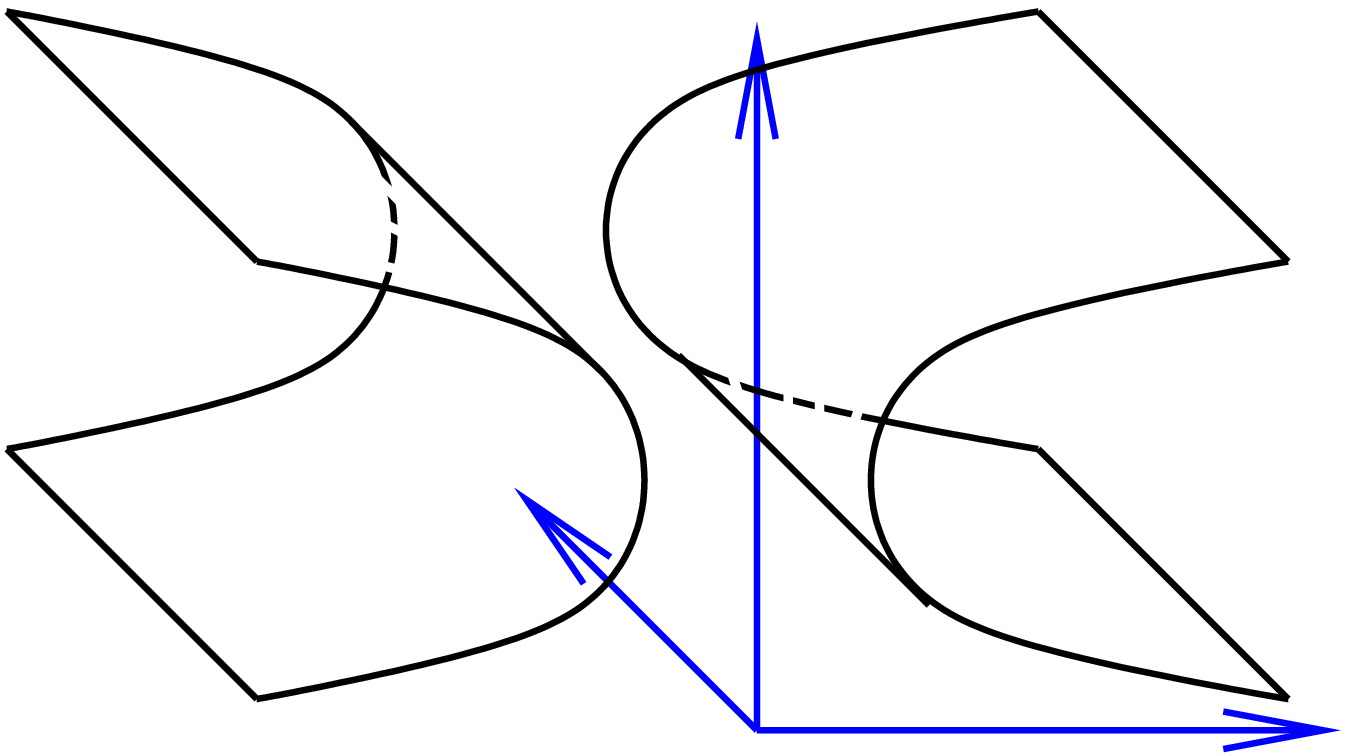}
\put(-90,90){$X^0(r)$}
\put(-124,38){$x_{p+1}$}
\put(-10,0){$r$}
\caption{A partial formation of a string while it decays away.}
\label{spike3}
\end{center}
\end{figure}

In addition to the above BIons, it is possible to find solutions
which are not infinitely extended in space or time.  By decreasing
the ratio of charge to mass in (\ref{bion-eq}), the solution
becomes a D-brane and anti-D-brane pair connected by a
throat.\cite{CM} Although the solution is not supersymmetric it is
static; similar solutions representing the annihilation of
D-branes and anti-D-branes have also been found.\cite{Savvidy} Our
intersecting S-brane has a similar deformation which is given by

\begin{equation} X^0=\int\frac{b_p  \ dr }{\sqrt{r^{2p-2}-r_0^{2p-2}}},
\hspace{.3in} A_{p+1}= \int \frac{a_p \
dr}{\sqrt{r^{2p-2}-r_0^{2p-2}}} \end{equation} where
$r_0^{2p-2}=a_p^2-b_p^2$. See Fig.~\ref{spike3}.
This S-brane has an interpretation as a partial formation of a
string.  There is a local fluctuation away from the vacuum which
tries but does not succeed in properly interpolating between
different vacua.  All of these solutions undergo worldvolume
signature change from spacelike to timelike worldvolume for
$r_0<r<b_p^{1/(p-1)}$.   The local fluctuation only exists for a
finite amount of time before disappearing.

\begin{figure}[htp]
\begin{minipage}{60mm}
\begin{center}
\includegraphics[width=6cm]{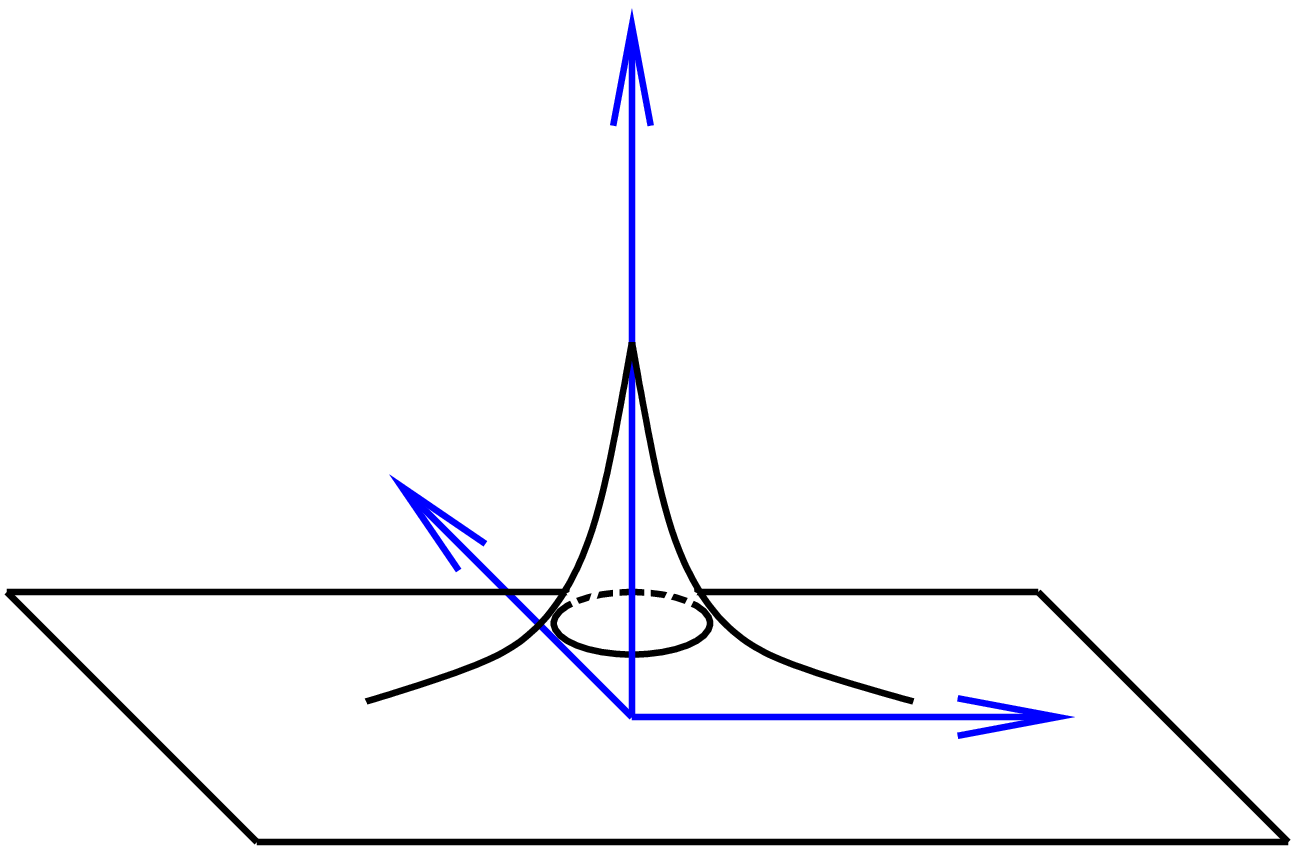}
\put(-40,24){$r$}
\put(-80,100){$\phi$}
\caption{A static BIon with a cusp at the top.}
\label{spike5}
\end{center}
\end{minipage}
\hspace*{2mm}
\begin{minipage}{60mm}
\begin{center}
\includegraphics[width=6cm]{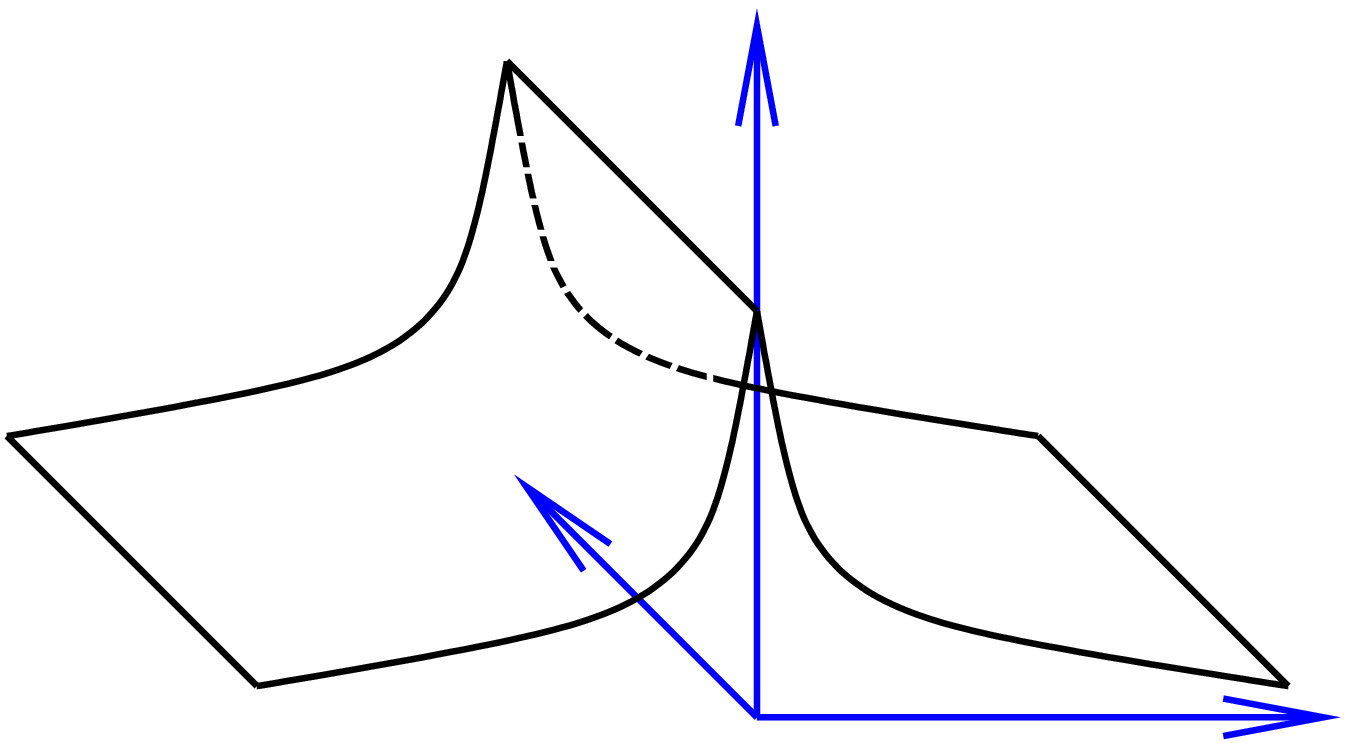}
\put(-70,90){$X^0(r)$}
\put(-118,38){$x_{p+1}$}
\put(-10,10){$r$}
\caption{A caustic is formed at a finite time, in a solution of the
 S-brane action.}
\label{spike4}
\end{center}
\end{minipage}
\end{figure}

While these S-brane solutions can be smooth, the solutions of
Refs.~\refcite{Sencre,LNT} describing the creation of co-dimension
one branes contain singularities.  Let us examine under what
conditions the D-brane and S-brane worldvolumes are smooth and
when they contain caustics and singularities. The D-brane and
anti-D-brane pair for example satisfy a constraint
$r_0^{2p-2}=b_p^2-a_p^2$ relating the radius of the throat $r_0$,
the height of the throat which is determined by $b_p$ and the
electric flux $a_p$. For the parameter range $b_p>a_p$, which can
be thought of as a subextremal limit, the throat is smooth. The
superextremal solutions $b_p<a_p$ describe one D-brane which comes
together forming a caustic\cite{CM,AH} past which we do not
properly understand what should happen (Fig.~\ref{spike5}). On the
other hand S-branes obey the constraint $r_0^{2p-2}=a_p^2-b_p^2$
so smooth solutions, $a_p>b_p$, are those where the field strength
is superextremal and larger than the timelike fluctuation.
Singular S-brane configurations are subextremal with $a_p<b_p$
(see Fig.~\ref{spike4}). Given the existence of smooth S-brane
solutions, however, one hopes and expects that related
non-singular solutions will be found using approaches similar to
that in Ref.~\refcite{Sencre,LNT}.

\section{Discussions}

By interpreting the role of Euclidean D-brane solutions during
brane decay and also explicit derivation from tachyon actions, we
have shown that the S-brane actions, first used and applied in
Refs.~\refcite{Sbraneaction} and \refcite{Sbraneevolution}, allow
for solutions describing the emergence of some closed string
physics.  In view of the fact that D-brane actions are fruitfully
applied for many results in string theory, we are potentially
still only scratching the surface, and it would be interesting to
see what further role S-brane actions have in physics.  The
equations of motion coming from the S-brane action are the same as
those from the D-brane action, thus at some level they are
basically the same. Conceptually, however, the S-brane actions are
different and so they potentially open up new directions for time
dependent D-brane dynamics.  Due to its worldvolume
parametrization, for example, S-branes can simultaneously describe
spatially disconnected objects.
We also emphasize that S-branes are not instantons due to the
background metric and fluctuations in the time direction. This
makes it possible to describe with ease the formation/annihilation
of D-branes/strings as seen in this review; it would be
interesting to consider possible connections to string production
during inflation.\cite{cosmic-string}   Since D-branes helped give
rise to many developments in string theory, such as AdS/CFT,
holography, Matrix theory, braneworlds and many other topics,
S-branes potentially may bring more than just generalizations of
the previous topics.  S-brane actions stand on the same footing as
the D-brane actions, and so we expect the possibility for their
use in many as yet undiscovered applications.

The $\beta$-function approach of string sigma models gives the
target space equations of motion, and in this sense the S-brane
actions are qualified to work at the same level as the usual
D-brane actions. But, as long as the S-brane action is intimately
related to the D-brane decay, it is important to clarify the
relation to the picture of time dependent tachyon condensation and
its deformation. The derivation of the action given in
Ref.~\refcite{Sbraneaction} was based on a homogeneous rolling
tachyon, while the S-brane solutions considered in this review
have large deformations from the homogeneous configuration. Thus
it is not straightforward to construct corresponding tachyon
solutions, as opposed to the situation of the static
BIons.\cite{Hirano} A subtle problem here is that there is no
``tachyon effective theory'' since the tachyon already has a mass
squared of order string scale. This suggests that one has to use
string field theories to describe inhomogeneous tachyon
condensation which in general appears quite difficult. Our S-brane
approach, we believe, may extract an important part of the
dynamics in a consistent manner with the string field theories. We
also note that the S-brane approach may be applied to various
field theories with topological defects, not only string theory,
because the derivation given in Ref.~\refcite{Sbraneaction} is
quite general. Applications along this direction should be
intriguing.

Although we have discussed the open string theory degrees of
freedom during tachyon condensation, there is no known decoupling
limit of closed string modes in this case.  In fact it is an
important and unresolved question as to what exactly is the
endstate of tachyon condensation called tachyon matter. In
Ref.~\refcite{cstring-decay} it was argued that this matter is
unstable and that the energy in the tachyon condensation process
will be transferred into very massive closed strings which then
disperse into the bulk.  Alternately Sen argues that while the
energy is transferred into closed strings, their dispersal is slow
and in fact tachyon matter is a specific form of stable closed
strings which can be described by open strings.  From this
viewpoint there may be a possible new open closed string duality
where a given open string background has a closed string
dual.\cite{sendual} An example of where this proposal has been
checked is in the $c=1$ matrix models.  Also, in trying to study
effects of gravitational backreaction on the brane decay process,
candidate dual supergravity solutions have also been constructed
in Ref.~\refcite{grav-duals} where it was found that their decay
time could be infinitely extended through explicit scaling limits;
insight into the possibility of a new open closed string duality
could arise from a careful study of their near horizon limit.
Recently Ref.~\refcite{Gut-Yi} has also argued that electric
fields act as a catalyst for the decay of unstable D1-branes decay
into closed strings and further increase the stability of the
tachyon condensate.  In general though the endpoint of tachyon
condensation is still a problem needing investigation.

With the inclusion of gauge fields on the D-brane worldvolume,
even if tachyon matter is unstable, however, the production of
closed strings is still at the heart of these S-brane solutions.

\section*{Acknowledgments}

K.H.~would like to thank his collaborators concerning this
subject, G.~W.~Gibbons, S.~Nagaoka, S.~Terashima, and P.~Yi. The
work by K.H. was supported in part by the Grant-in-Aid for
Scientific Research (No.~12440060, 13135205, 15540256 and
15740143) from the Japan Ministry of Education, Science and
Culture.  P.M.H. and J.E.W  are supported in part by the National
Science Council, the Center for Theoretical Physics at National
Taiwan University and the National Center for Theoretical
Sciences.


\end{document}